\documentclass{ws-procs961x669} 
\def\be{\begin{equation}}
\def\ee{\end{equation}}
\def\ba{\begin{eqnarray}}
\def\ea{\end{eqnarray}}

\def\lb{\label}

 \def\fL{\lambda}
 \def\la{\lambda}

\def\R{{\hat R}}
\def\X{{\hat g}}
\def\hnabla{\hat{\nabla}}
\def\hGamma{\hat{\Gamma}}
\begin{document}

\title{Palatini kinetic scalar-tensor theory: analytical and numerical solutions}
\author{  D. V. Gal'tsov$^{*}$ and D. S.  Bushuev$^{**}$}

\address{Faculty of Physics, Moscow State University, 119899, Moscow, Russia\\
$^{*}$E-mail: galtsov@phys.msu.ru\\
$^{**}$E-mail:bushuev.ds16@physics.msu.ru }

\begin{abstract}
We investigate static spherically symmetric solutions in the Palatini kinetically coupled scalar-tensor theory, which reduces to gravity minimally coupled to a scalar field in Einstein frame. Using the fact that the Jordan and Einstein frame  are related by a reversible disformal transformation, which can be solved in  closed form, we derive the general solution in the Jordan frame and show that it does not contain black holes. There is a wormhole branch and a naked singularity branch between which  lies a non-singular asymptotically flat solution with kernel $M_{1,1}\times S^2$. This theory strongly violates the null energy condition.
  
\end{abstract}


\bodymatter
\section{Introduction}
Recently~\cite{Galtsov:2018xuc} was discovered a particularly simple, albeit nontrivial, example of scalar-tensor theory with kinetic coupling, which strongly violates the null energy condition.  
This theory is a Palatini version of the well-known model with a scalar field  coupled to Ricci tensor and Ricci scalar through derivatives \cite{Amendola:1993uh}. With a  certain ratio of the two coupling constants, the metric version of this theory belongs to Horndeski class~\cite{Horndeski,Kobayashi:2019hrl,Deffayet:2011gz}, and has an attractive feature to generate inflationary cosmological solutions without scalar potentials~\cite{Capozziello:1999uwa, Sushkov:2009hk,Granda:2010hb,Granda:2017dlx}. 

The situation is   different in the Palatini version of this theory, what is typical for scalar-tensor theories in general~ \cite{Sotiriou:2005hu,Harko:2010hw, Luo:2014eda,Helpin:2019kcq,Gumjudpai:2016ioy}. 
In this case, another ratio of the two coupling constants
gives a theory admitting an  Einstein frame  in which the scalar coupling is minimally coupled. Moreover, the Einstein metric turns out to be the metric, the Levi-Civita connection of which is the Palatini connection of the original theory.  The two metrics  are related by an invertible disformal transformation~ \cite{Bekenstein:1992pj,Bettoni:2013diz,Sakstein:2015jca}. The reversibilty of the disformal transformations ensures  classical equivalence of the theories in two frames~ \cite{Domenech:2015tca,Takahashi:2017zgr,Babichev:2019twf}, which probably extends to the quantum level~ \cite{Kamenshchik:2014waa}.  It also opens a way to use disformal transformations as a generating tool to construct exact solutions of new theories~\cite{Galtsov:2018xuc,BenAchour:2020wiw}. Some examples of  solutions of Palatini kinetic theory were presented in ~\cite{Galtsov:2018xuc} demonstrating desingularization of solutions which are singular in the Einstein frame. 
For deeper discussion of singularities of general relativity in the context of modified gravity theories see~\cite{Fewster:2010gm,Fewster:2019bjg}.
 
Further investigation of~\cite{Galtsov:2020jnu} showed  a connection between this  theory   and the metric conformally coupled theory, both of which demonstrate a strong violation of the null energy condition. The natural question is whether our theory allows for wormhole solutions ~\cite{Rubakov:2014jja,Mandal2019}. Here we show that this is indeed the case. 
 
\section{The setup}
We consider the action with two independent couplings of the derivatives of the scalar field  $\phi_{\mu}\equiv\phi_{,\mu}$ to the Ricci tensor and the Ricci scalar\cite{Amendola:1993uh,Capozziello:1999uwa,Sushkov:2009hk,Granda:2010hb} 
\begin{equation}
  S  = \int   d^{4}x\sqrt{-g}\left[R  -  \left(g_{\mu\nu}  +  \kappa_{1} g_{\mu\nu}R  +  \kappa_{2}R_{\mu\nu}\right)\phi^\mu\phi^\nu\right],
\end{equation}
where $\phi^\mu=\phi_\nu g^{\mu\nu}$. 
In the metric formalism, an interesting case is $\kappa_1=-\kappa_2/2$ when the Ricci tensor and Ricci scalar are combined into the Einstein tensor, which has zero covariant divergence. This case belongs to the Horndeski ghost-free class. 

In the Palatini version, the action will read 
\begin{align}\lb{kP}
  S =\int d^{4}x\sqrt{-g}   
\big[(\R_{\mu\nu}-\phi_\mu\phi_\nu) g^{\mu\nu}-
\R_{\alpha\beta} \phi_\mu\phi_\nu (\kappa_1   g^{\alpha\beta}g^{\mu\nu}  +\kappa_{2}   g^{\alpha\mu}g^{\beta\nu})\big],
\end{align}
where $\R_{\alpha\beta}$ is the Ricci-tensor constructed with the Palatini  connection $\hGamma$. The corresponding Einstein equations are 
\be \lb{PaE}
\fL\R_{\mu\nu}-\phi_\mu\phi_\nu(1+\kappa_1 \R)  -2\kappa_2 \R_{\alpha(\mu}\phi_{\nu)}\phi^\alpha- g_{\mu\nu}  L/2 = 0,
\ee
where the Lagrangian and the factor $\la$ are
\be\lb{LZ}
L=\R_{\mu\nu}Z^{\mu\nu}-\phi_\mu\phi_\nu g^{\mu\nu},\;\; Z^{\mu\nu}
  =\fL g^{\mu\nu} - \kappa_2\phi^\mu\phi^\nu,\;\;\fL = (1 - \kappa_1 X),   
\ee
and  $X= \phi_\alpha\phi_\beta g^{\alpha\beta}$. Variation over $\phi$ gives rise to a scalar equation
\be\lb{sp}
\partial_\mu\left[ \sqrt{-g} \left( \phi^\mu+\kappa_1 \R\phi^\mu+\kappa_2
\R_{\alpha\beta}g^{\beta\mu}\phi^\alpha\right)\right]=0,
\ee
and variation over an independent connection gives:
\be
\lb{eqZ}
  \hnabla_\lambda \left( \sqrt{-g}Z^{\mu\nu}\right) = 0,
\ee
where the covariant derivative refers to the $\hGamma$-connection.  

Direct solution of the Einstein-scalar equations (\ref{PaE},\ref{sp}) does not seem possible. But as was shown in \cite{Galtsov:2018xuc}, the Palatini equation (\ref{eqZ}) can be solved giving the Levi-Civita connection  
 \be
 \hGamma^\lambda_{\mu\nu}= \X^{\la\tau}\left(\partial_\mu \X_{\la\nu}+ \partial_\mu\X_{\mu\la}-\partial_\la \X_{\mu\nu}\right)/2
 \ee
of the second metric 
\be\lb{dis}
{\hat g}_{\mu\nu}=\sqrt{\Lambda\lambda}\left(g_{\mu\nu}+\kappa_2 \Lambda^{-1} \phi_\mu \phi_\nu  \right),\quad 
 \Lambda=1 -(\kappa_1+\kappa_2)X .\ee 
This is a disformal transformation, which can be inverted (for details see~\cite{Galtsov:2018xuc}).  It is clear that setting $\kappa_1+\kappa_2=0$, will greatly simplify the above relation, and we assume that from now on.  
Then, remarkably, when expressed in the second metric, the action (\ref{kP}) is transformed into an Einstein-Hilbert action plus a minimally coupled scalar: 
 \be\lb{MES}
S_E=\int \sqrt{-\X}\left[R_{\mu\nu}(\X)-\phi_\mu\phi_\nu\right]\X^{\mu\nu} d^4x.
 \ee
This action gives (setting torsion to zero) the same equations in the Palatini version and in the metric version: 
 \be\lb{EX}
 R_{\mu\nu}=\phi_\mu\phi_\nu,\qquad {\hat\Box}\phi=0.
 \ee
So our strategy will be first to integrate the system  (\ref{EX}) and then to  use an inverse of the disformal transformation (\ref{dis}) to obtain the desired metric $g_{\mu\nu}$. To be sure that we get the general solutions in the Jordan frame we need a general solution in the Einstein frame. 
\section{Einstein frame: complete integration}
Static solutions of the minimally coupled scalar-tensor gravity were investigated in the past  in various contexts ~\cite{Fisher,Bergmann:1957zza,Bronnikov:1973fh,JNW}. The most popular solution belongs to Fisher~\cite{Fisher}.  To make sure that it is unique in the class of static spherically symmetric solutions, we repeat the  derivation here. 

It is convenient to present the metric ansatz as
\be
	ds^2=Adt^2+\frac{dr^2}{A}+R(d\theta^2+\sin^2{\theta}d\phi^2),
\ee
 where $A$ and $R$ are functions of $r$ only, and it is assumed that $R>0$ (note that we are using $R$ and not $R^2$ in the angular part).
The non-zero components of the Ricci tensor are
\begin{align}
	&R_{tt}=\frac{A}{2R}(A^\prime R)^\prime,\\
	&R_{rr}=-\frac{1}{2A R^2}(2ARR^{\prime \prime}+A^\prime R R^\prime-A {R^\prime}^2+A^{\prime \prime}R^2),\\
	&R_{\theta\theta}=\frac{1}{\sin^2{\theta}}R_{\phi\phi}=-\frac{1}{2}(AR^\prime)^\prime+1,
\end{align}
where the prime denotes the radial derivative. The scalar field satisfies the simple divergence equation 
\be
	(AR\phi^\prime)^\prime=0.\ee
With our choice of the action, the Einstein equations read
	$R_{\mu\nu}=\phi_\mu\phi_\nu,$
	so the equations for $(tt)$ and $(\theta\theta)$ components of the Ricci tensor are sourceless
	 and can be integrated once, giving
	$$A^\prime  R=C={\rm const}; \quad
	AR^\prime=2(r+a),\quad a={\rm const}.$$
	Similarly, integrating the scalar equation   we get
$$
	\phi^\prime=\frac{q}{AR}, 
$$
where the integration constant  $q $ represents the scalar charge.
	
From the $(rr)$ component of the Einstein equation we obtain a constraint
	$$
	(2R-CR^\prime)A-2(r+a)^2+q^2=0,
$$
Finally,  we derive   a separate equation for $R$:
$$
	\frac{2}{R^\prime}-\frac{2(r+a)R^{\prime\prime}}{{R^\prime}^2}=\frac{C}{R}.
$$
To solve it, we put
	$R=e^u$,
	obtaining a non-linear equation for u:
$$
	u^{\prime\prime}+{u^\prime}^2-\frac{u^\prime}{x}+\frac{C{u^\prime}^2}{2x}=0, 
$$
where $x=r+a.$
	This equation can be integrated once, giving
$$
	u^\prime=\frac{2x}{x^2+Cx+2C_1}
$$
	with some new integration constant $C_1$. This constant, using the constraint equation,  is reduced to 
$
	C_1=-q^2/4.
$
Further integration gives
$$
	u=\ln{(2x^2+2Cx-q^2)}+\frac{2C}{\sqrt{C^2+2q^2}}\tanh^{-1}{\frac{2x+C}{\sqrt{C^2+2q^2}}}+u_0,
$$
	where $u_0$ is a trivial integration constant which can be changed by angular coordinates rescaling, so we put $u_0=-\ln{2}$. Rewriting the quadratic function under logarithm as
	$$
	x^2+Cx-\frac{q^2}{2}=(x-x_+)(x-x_-),\quad x_\pm=-\frac{1}{2}\left(C\pm\sqrt{C^2+2q^2}\right).
$$ 
To get the metric function $R$, one has to take the power that leads to:
\be\lb{Ru}
	R=e^u=(x-x_-)(x-x_+)\left(\frac{x-x_-}{x-x_+}\right)^\gamma=r^2\left(1-\frac{b}{r}\right)^{1-\gamma},
\ee
	where we have introduced two new parameters instead of $C$, $q^2$:
\be
	\gamma=\frac{C}{\sqrt{C^2+2q^2}}\quad b=\sqrt{C^2+2q^2},
\ee
and a new radial coordinate $r=x-x_-$.
The metric function $A$ in the new variables reads
$$
	A=\frac{2(r+x_-)}{R^\prime}.
$$
	Differentiating (\ref{Ru}), we find
\be
	A=\left(1-\frac{b}{r}\right)^\gamma.
\ee
Summarizing, we obtained the famous Fisher-Janis-Newman-Winicour (FJNW) solution, proving its uniqueness within the static spherically symmetric class:
\be
	ds^2=-\left(1-\frac{b}{r}\right)^\gamma dt^2+\left(1-\frac{b}{r}\right)^{-\gamma}dr^2+r^2\left(1-\frac{b}{r}\right)^{1-\gamma}d\Omega.
\ee
The derivative of the scalar field in terms of $b,\,\gamma$ reads:
\be \phi'=\frac{b\sqrt{1-\gamma^2}}{\sqrt{2}\,r(r-b)}
\ee
Note that the range of the radial variable is chosen $r\in(b,\infty)$ to ensure positivity of the function $R$. The region $r<b$ corresponds to cosmological solution\cite{Abdolrahimi:2009dc}.	
\section{Disformal transformation}
	To pass to the Jordan frame, it is necessary to carry out  the inverse disformal transformation 
\be
	g_{\mu\nu}=\hat{g}_{\mu\nu}\lambda^{-1/2}-\kappa\phi_\mu\phi_\nu,
\ee
	where
\be
	\lambda=\left(g_{rr}/w\right)^{-2/3},\quad w=A=\left(1-\frac{b}{r}\right)^{-\gamma},
\ee
	and $g_{rr}$ obeys the following equation:
	\be
\left(g_{rr}-\frac{2x}{3\sqrt{3}}\right)^3=w g_{rr},\quad x=\frac{3\sqrt{3}\kappa q^2}{2 r^2(r-b)^2}.
\ee
	A real solution of this equation reads
\be
	g_{rr}=\frac{2x}{3\sqrt{3}}+\frac{1}{\sqrt{3}}\begin{cases}
	2w\cos\left({\frac{1}{3}\arccos{x/w}}\right),& x<w,\\w^{2/3} B+w^{1/3}B^{-1},& x>w,
	\end{cases}
\ee
	where
$$
	B=\left(x+\sqrt{x^2-w^2}\right)^{1/3}.
$$
	Two other Jordan frame metric components are
\be
	g_{tt}=\hat{g}_{tt}\lambda^{-1/2}\qquad g_{\theta\theta}\equiv Q=\hat{g}_{\theta\theta}\lambda^{-1/2}=R\lambda^{-1/2}.
\ee
 
	It is easy to see that the above transformation to the Jordan frame preserves asymptotic flatness. Indeed, when $r\rightarrow \infty$, the variables $x\rightarrow 0$, $w\rightarrow 1$, so that
	$$
	g_{rr}\sim 1+x/\sqrt{3}\Rightarrow \lambda=1+O(1/r^4),
$$
	and we get as $r\rightarrow \infty$:
\be
	g_{tt}\simeq -1+\frac{\gamma b}{r}\quad g_{rr}\simeq 1-\frac{\gamma b}{r}\quad g_{\theta\theta}\sim r^2.
\ee
\section{Wormholes}
Consider first an asymptotic behavior of the Jordan metric as $r\rightarrow b$. Without loss of generality, let the dimensionless parameter $\kappa q^2/b^4 $ be equal to one. Then a careful analysis of the behavior of various terms for $x\rightarrow\infty,\,w\rightarrow 0$ assuming arbitrary   values $0<\gamma<1$ gives the following leading approximation for small $u=(r/b-1)$: 
\be
	ds^2=-  u^{2(2\gamma-1)/3}dt^2+\frac{  b^2 du^2}{u^2}+  b^2 u^{(1-2\gamma)/3}
	d\Omega.
	\ee
	As $r\rightarrow b$, the metric component  $g_{\theta\theta}=Q(r)$ explodes  if $\gamma>1/2$, while it goes to zero for $\gamma<1/2$.  One can suspect that in the first case the intrinsic volume of a region   close to $r=b$, namely $r\in [b+\delta,\,r_0]$, such that $\delta\ll r_0-b\ll b$,  will be infinite as $\delta\to 0$.
	Indeed,
$$
	V_\delta=4\pi\int_{b+\delta}^{r_0}{\sqrt{h}dr }= \frac{12\pi }{2\gamma-1} b^3\left(\frac{\delta}{b}\right)^{(1-2\gamma)/3}.
$$
Thus, for $\gamma>1/2$ one has the second infinite region of three-space. This region is not flat, and not Ricci flat. It is not asymptotically flat as $r\to b$ either, moreover the scalar field diverges there. But one can suspect the existence of an asymmetric wormhole, connecting our AF-region with the second sheet to the left of the 
  local minimum of $Q(r)$ (a turning point) at some point $r_t>b$, if exists: 
\be
	\frac{dQ}{dr}\bigg|_{r=r_t}=0,\quad\frac{d^2Q}{dr^2}\bigg|_{r=r_t}>0,
\ee 
which could correspond to a wormhole throat. 
	 
	Numerical analysis shows that for $\gamma<1/2$, $g_{\theta\theta}$ does not have a local minimum for all positive values of the coupling constant $\kappa$ . For $\gamma>1/2$ a local minimum arises, with $r_t$ depending on $\gamma$ and $\kappa$.
 \subsection{Numerical solutions}	
Though we obtained solutions in the Jordan frame analytically, the algebraic functions involved are difficult for understanding the nature of the metric, so we undertake numerical plotting. On Fig. 1, 2 we plot the metric functions for various values of $1/2<\gamma<1$. In this case a local minimum of $Q(r)$ arises at   $r=r_t$,  whose position depends on $\gamma$ and $\kappa$.  This corresponds to the throat radius of an asymmetric wormhole. The values $r_t$  can be found numerically with any desired accuracy.  
For $\kappa=2$,  
the radii of the throat for some selected $\gamma$ are	\begin{center}\begin{tabular}{c|c
	|c|c|c}
		$\gamma$ &3/5& 3/4 &0,95 &0,99 \\\hline
		$r_t/b$&1.041&1.078&1.072&1.041 \end{tabular}
\end{center}
	With growing $\kappa$, the throat radius increases, as can be seen from the left panel of Fig.1.
	\begin{figure}[h]\lb{Fig. 1}
	\center{\includegraphics[scale=0.5]{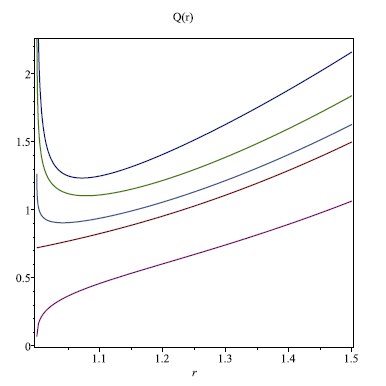}\includegraphics[scale=0.5]{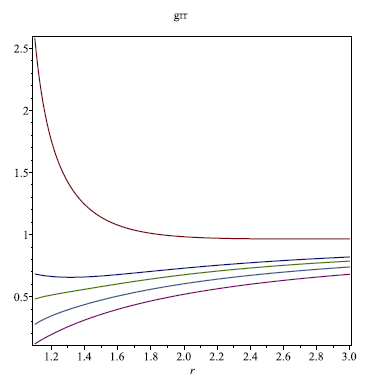}}
	\caption{Metric functions $g_{\theta\theta}$ and $g_{tt}$ for various values of   $\gamma$: $\gamma=0.1$ (purple), $\gamma=0.5$ (red), $\gamma=0.6$ (light blue), $\gamma=0.75$ (green), $\gamma=0.95$ (blue).} \label{fig:metricQrr}
	\end{figure}
			\begin{figure}[h]
	\center{\includegraphics[scale=0.5]{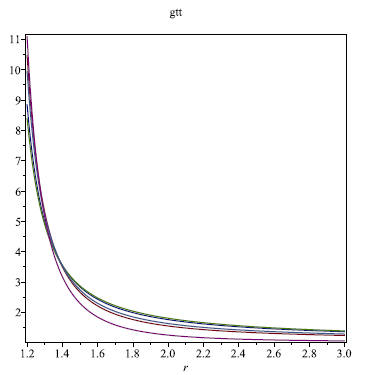}\includegraphics[scale=0.5]{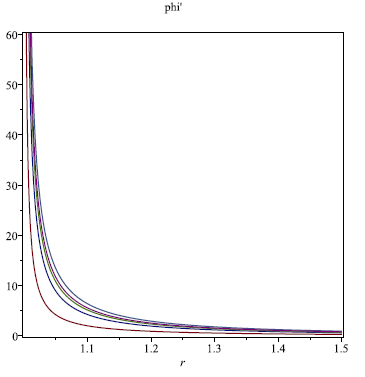}}
	\caption{Metric function $g_{rr}$ and the derivative of the scalar field ($\phi^\prime$) for various values of  $\gamma:\;$ $\gamma=0.1$ (purple), $\gamma=0.5$ (red), $\gamma=0.6$ (light blue), $\gamma=0.75$ (green), $\gamma=0.95$ (blue).} \label{fig:metricttphi}
	\end{figure}
	 \begin{figure}[h]
	\center{\includegraphics[scale=0.34]{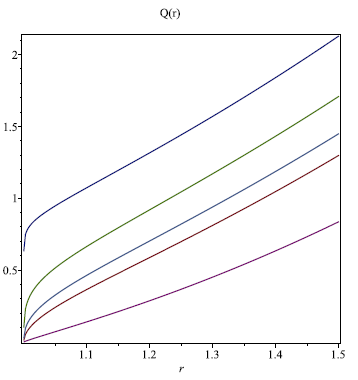}\includegraphics[scale=0.34]{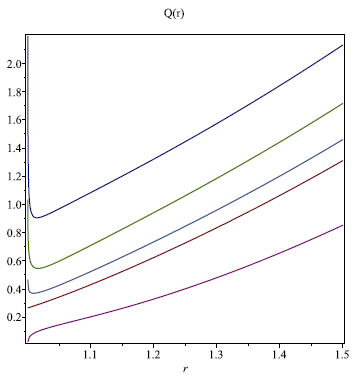}\includegraphics[scale=0.34]{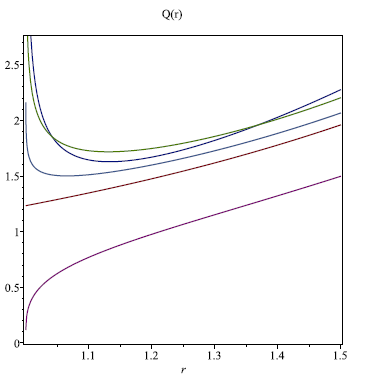}}
	\caption{Metric function ($g_{\theta\theta}$) for various values of   $\kappa$: $\kappa=0$ (left figure), $\kappa=0.01$ (center), $\kappa=10$ (right); and $\gamma$: $\gamma=0.1$ (purple), $\gamma=0.5$ (red), $\gamma=0.6$ (light blue), $\gamma=0.75$ (green), $\gamma=0.95$ (blue).} \label{fig:metricQ}
	\end{figure}	
Left to the minimum, the radial function $Q$ increases again and tends to infinity as $r\to b$ for $\gamma>1/2$. 
 From Fig. 2 (left panel) one can also see that $g_{tt}>0$ for all $r>b$, so the solutions do not have horizons. 
 
 It can be shown that our wormhole is traversable: radial time-like and null geodesics freely pass through the throat. The tidal forces remain finite.
 \subsection{Geometry of the second sheet }
 To investigate the geometry of the second leaf of the wormhole $ r <r_t $, we first calculate the asymptotic behavior of the Ricci tensor and the Ricci scalar constructed from the Jordan metric (which is not the Palatini Ricci tensor): 
\begin{align} 
	&R_{\mu\nu}(g)dx^\mu dx^\nu=-\frac{(1-2\gamma)^2}{6u^2}du^2+d\Omega,\\
	&R(g)=\frac{2u^{(2\gamma-1)/3}}{b^2}-\frac{(1-2\gamma)^2}{6\mu^2}.
\end{align}
The $rr$-component of the Ricci tensor diverges as $u\to 0$, but the square of the Ricci tensor is finite for $\gamma>1/2$:
\be
R_{\mu\nu}R^{\mu\nu}=\frac{2 u^{2(2\gamma-1)/3}}{b^4}+\frac{(2\gamma-1)^2}{36\, b^4}.
 \ee

 We then calculate the effective energy density and pressure by defining them in terms of the mixed components of the (diagonal) Einstein tensor, again constructed using the Jordan metric: 
 \be
 G_\mu^\nu(g)=(-\epsilon,\; p_r,\; p_\theta,\; p_\varphi).
 \ee
 This is not the physical energy density and pressure in Palatini's theory (see below), but only the effective ones used to illustrate geometry in terms of the notions of the Riemann geometry. The calculation gives for the energy density and radial pressure: 
\be \epsilon =\frac{ u^{(2\gamma-1)/3}}{b^2}-\frac{(2\gamma-1)^2}{12b^2},\qquad p_r=-\frac{u^{(2\gamma-1)/3}}{b^2}-\frac{(2\gamma-1)^2}{12b^2}.\ee
Their sum is negative, demonstrating violation of NEC in our theory. The tangential pressure is positive and finite
\be p_\theta=p_\varphi=(2\gamma-1)^2/{12b^2}.\ee

In the sense of the Palatini theory, the Ricci scalar must be computed by contracting the Ricci tensor built from the Palatini connection, which is the Levi-Civita connection of  Einstein's frame  (i.e.     $R_{\mu\nu}(\X)$),  with  the Jordan contravariant metric tensor. Near the boundary $r=b$ the leading behavior of this quantity will be:
\be
R_{P}=R_{\mu\nu}(\X)\,g^{\mu\nu}=\frac{(1-2\gamma)^2}{6\, b^2}\frac1{
u^{2-\gamma}},
\ee 
which diverges as $u\to 0$  for any $0<\gamma<1$.   The question arises whether an observer will see the second sheet as a regular or a singular space. In fact, with  independent connection, the curves of   minimum length do not coincide with the curves along which the tangent vector is   parallel transported. More generally, the matter action can depend only on the metric, or on the metric and the connection. The point particle action belongs to the first type, so the particle will move along the  curves of minimal length. Such an observer will not see  singularities on the internal sheet of the wormhole. If the notion of geodesic completeness is attributed to curves of minimal length, then this spacetime should be regarded as nonsingular indeed. 
\subsection{Isotropic coordinates}
 The asymptotic metric of the second sheet can be presented in isotropic coordinates $t,\rho,\theta,\phi$ changing the radial variable as
$$
	\rho=b\exp{\left(\nu^{-1}u^\nu\right)},\quad  \nu= {(2\gamma-1)}/{6 }.$$
The resulting metric reads	\be ds^2=-\left(\nu\ln\frac{\rho}{b}\right)^4 dt^2+ f^2 dl_3^2, \quad dl_3^2=d\rho^2+\rho^2 d\Omega.
\ee The conformal factor   $f =\frac{b}{\rho\nu} \left(\ln{\rho/b}\right)^{-1}$, shown in Fig.4 as a function of $\rho$, clearly demonstrates two asymptotic behavior of a wormhole:
 	\begin{figure}[h]
 	\center{\includegraphics[scale=.35
 	]{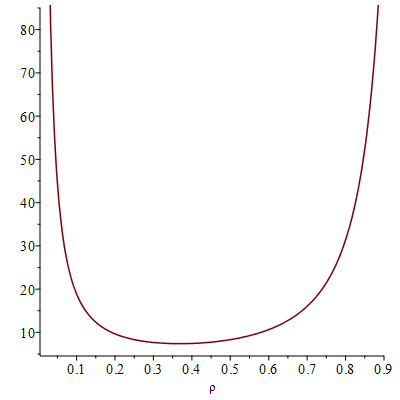}}
	 \caption{The radial scale factor in isotropic coordinates.} \label{fig:f2}
 	\end{figure}
 	\subsection{ The soliton $\gamma=1/2$}
In the Einstein frame the FJNW solution for $\gamma=1/2$ satisfies the condition $g_{rr}=1/|g_{tt}|$:
\be
	ds^2=-\sqrt{1-\frac{b}{r}} dt^2+ \frac{dr^2}{\sqrt{1-\frac{b}{r}}}+r^2\sqrt{1-\frac{b}{r}}d\Omega,
\ee
It has  curvature invariants divergent as $r\to b$. But its Jordan frame counterpart turns out to be non-singular, namely
\be
  ds^2=-dt^2+\xi^{-2} dr^2 +b^2(d\theta^2+\sin^2\theta d\varphi^2).
\ee
Passing to a new radial coordinate $z=b\ln \xi$, we transform  the domain $ r\in (b, \infty)$ into the complete real line $-\infty<z<\infty$
\be
  ds^2=-dt^2+dz^2 +b^2(d\theta^2+\sin^2\theta d\varphi^2).
\ee
Therefore the manifold is isomorphic to the product of the  two-dimensional Minkowski space and a sphere of  radius $b$: $M_{1,1}\times S^2$.  This manifold is geodesically complete in the Riemannian geometry sense. Thus, our solution can be regarded as a  regular scalar-tensor soliton. Its striking feature is that, within the framework of our theory, it is supported by a singular scalar. The singularity now is transfered to the Palatini connection. 
\subsection{Naked singularities}
The range of the power index  $0<\gamma<1/2$	corresponds to Jordan metrics singular in the Riemannian sense.	This can be  seen already from inspection of the Ricci scalar constructed with the Jordan metric: 
\be R(g)=\frac{2u^{(2\gamma-1)/3}}{b^2}-\frac{(1-2\gamma)^2}{6b^2}.
\ee
The constant term is due to the two-sphere sector of the metris, while the first term originates from radial component of the Ricci tensor and it tends to infinity for $\gamma<1/2$.

Also, one can see that the three-volume of the region between $r=b-\delta$ and $r=b$ for $\gamma<1/2$ tends to zero
when $\delta\to 0$:
\be
	V_\delta=  12\pi  b^3\left( {\delta}/{b}\right)^{(1-2\gamma)/3}.
\ee
The metric function $g_{\theta\theta}$ for various $\kappa$ and $\gamma$ is plotted in Fig.3.   
 From these plots  one can see that in all three cases the metric has no horizons. More general numerical tests show that  the static spherical sector of our theory does not  contain black holes indeed.

\subsection{Classification of solutions} 
Since our solution in the Einstein frame is generic in its class,  we conclude that the above three types of static spherically symmetric solutions in the Jordan frame exhaust all   possibilities, namely: 

	\begin{tabular}{r p{380pt}}
		$a)$ $\gamma>1/2$ --- &   asymmetric wormholes,  
	\\
		$b)$ $\gamma=1/2$ --- &   regular soliton   with the $M_{1,1}\times S_2$  core,\\
		$c)$ $\gamma<1/2$ --- & asymptotically flat spacetime with a naked sigularity.
	\end{tabular}
	No  solutions contain 	horizons.

Finally, we have two options for introducing the Kretschman scalar to investigate non-Ricci singularities.  The Riemannian type scalar is simply $K_R=  R_{\mu\nu\lambda\tau}(g)R^{\mu\nu\lambda\tau}(g).$
The Palatini-type Kretschmann scalar reads $$K_P= R^\mu_{\;\nu\lambda\tau}(\X)R^\alpha_{\;\beta\gamma\delta}(\X)\,g_{\mu\alpha}\,g^{\nu\beta}\, g^{\lambda\gamma}\,g^{\tau\delta}.$$
In the asymptotic region of the second sheet (small $u$) the first one is:
\be
	K_R=\frac{1}{9\,b^4}\left( 36u^{2(2\gamma-1)/3}-2(2\gamma-1)^2 u^{(2\gamma-1)/3}+12\gamma^2(\gamma-1)^2+6\gamma(\gamma-1)  +3/4\right).
\ee
It is finite in the   cases $a, b$  and singular in the case $c$ as expected. The  Palatini Kretschmann scalar always diverges as $u\to 0$:
\be
K_P=\frac{(\gamma-1)^2(5\gamma^2-2\gamma+2)}{4b^4\,u^{4(2-\gamma)/3}},
\ee
reflecting singular nature of the Palatini connection.
 
\section{Conclusion}
To summarize: we have found that Palatini scalar-tensor theory with a scalar field kinetically coupled to the Ricci tensor and the Ricci scalar is an example of theory that does not admit  solutions with an event horizon at least in the static case. 

The theory has Palatini connection which is the Levi-Civita connection of the Einstein frame metic. The general static spherically symmetric solution in the Einstein frame is given by the Fisher family of  metrics, specified by the power index $0<\gamma<1$. We have shown that  for $1/2<\gamma<1$ the Jordan metric describes wormholes, interpolating between an asymptotically flat space and a Ricci non-flat second space of an infinite volume.The second sheet of the wormholes is   not asymptotically flat. Its spacetime metric is regular, and the Riemannian Einstein tensor corresponds to an effective matter violating the null energy condition. 

For $\gamma<1/2$ the Jordan solutions are one-sheeted and represent naked singularities in  asymptotically flat space. In the intermediate case  $\gamma=1/2$ the Jordan metric is non-singular, and have a core of cylindrical topology.

  The work is supported by the Russian Foundation for Basic Research on the project 20-52-18012Bulg-a, and the Scientific and Educational School of Moscow State University “Fundamental and Applied Space Research”.


\begin{thebibliography}{4}
 \bibitem{Galtsov:2018xuc}
  D.~Gal'tsov and S.~Zhidkova,
  ``Ghost-free Palatini derivative scalar-tensor theory: Desingularization and the speed test,''
  Phys.\ Lett.\ B {\bf 790} (2019) 453
  [arXiv:1808.00492 [hep-th]].

\bibitem{Amendola:1993uh}
  L.~Amendola,
  ``Cosmology with nonminimal derivative couplings,''
  Phys.\ Lett.\ B {\bf 301}, 175 (1993).
\bibitem{Horndeski}  G. W. Horndeski,
``Second-order scalar-tensor field equations in a four-dimensional space,
Int. J. Theor. Phys. {\bf 10}, 363-384 (1974).

\bibitem{Kobayashi:2019hrl} 
  T.~Kobayashi,
  ``Horndeski theory and beyond: a review,''
  Rept.\ Prog.\ Phys.\  {\bf 82}, no. 8, 086901 (2019)
  [arXiv:1901.07183 [gr-qc]].
  
\bibitem{Deffayet:2011gz}
C.~Deffayet, X.~Gao, D.~A.~Steer and G.~Zahariade,
``From k-essence to generalised Galileons,''
Phys. Rev. D \textbf{84} (2011), 064039
[arXiv:1103.3260 [hep-th]].
 
\bibitem{Capozziello:1999uwa}
  S.~Capozziello and G.~Lambiase,
  ``Nonminimal derivative coupling and the recovering of cosmological constant,''
  Gen.\ Rel.\ Grav.\  {\bf 31} (1999) 1005
  [gr-qc/9901051].

\bibitem{Sushkov:2009hk}
S.~V.~Sushkov,
``Exact cosmological solutions with nonminimal derivative coupling,''
Phys. Rev. D \textbf{80} (2009), 103505
[arXiv:0910.0980 [gr-qc]].

  \bibitem{Granda:2010hb}
  L.~N.~Granda and W.~Cardona,
  ``General Non-minimal Kinetic coupling to gravity,''
  JCAP {\bf 1007}, 021 (2010).
  [arXiv:1005.2716 [hep-th]].

\bibitem{Granda:2017dlx}
  L.~N.~Granda and D.~F.~Jimenez,
  ``Dynamical analysis for a scalar-tensor model with kinetic and nonminimal couplings,''
  Int.\ J.\ Mod.\ Phys.\ D {\bf 27} (2017) no.03,  1850030
  [arXiv:1710.07273 [gr-qc]].
\bibitem{Sotiriou:2005hu}
  T.~P.~Sotiriou,
  ``Unification of inflation and cosmic acceleration in the Palatini formalism,''
  Phys.\ Rev.\ D {\bf 73}, 063515 (2006).
  [gr-qc/0509029].

  \bibitem{Harko:2010hw}
  T.~Harko, T.~S.~Koivisto and F.~S.~N.~Lobo,
  ``Palatini formulation of modified gravity with a nonminimal curvature-matter coupling,''
  Mod.\ Phys.\ Lett.\ A {\bf 26}, 1467 (2011).
  [arXiv:1007.4415 [gr-qc]].

     \bibitem{Luo:2014eda}
  X.~Luo, P.~Wu and H.~Yu,
  ``Non-minimal derivatively coupled quintessence in the Palatini formalism,''
  Astrophys.\ Space Sci.\  {\bf 350}, no. 2, 831 (2014).
  
\bibitem{Helpin:2019kcq}
  T.~Helpin and M.~S.~Volkov,
  ``Varying the Horndeski Lagrangian within the Palatini approach,''
  arXiv:1906.07607 [hep-th].

\bibitem{Gumjudpai:2016ioy}
  N.~Kaewkhao and B.~Gumjudpai,
  ``Cosmology of non-minimal derivative coupling to gravity in Palatini formalism and its chaotic inflation,''
  Phys.\ Dark Univ.\  {\bf 20}, 20 (2018).
  
  \bibitem{Bekenstein:1992pj}
  J.~D.~Bekenstein,
  ``The Relation between physical and gravitational geometry,''
  Phys.\ Rev.\ D {\bf 48}, 3641 (1993).
 [gr-qc/9211017].
 
 \bibitem{Bettoni:2013diz}
  D.~Bettoni and S.~Liberati,
  ``Disformal invariance of second order scalar-tensor theories: Framing the Horndeski action,''
  Phys.\ Rev.\ D {\bf 88}, 084020 (2013).
   [arXiv:1306.6724 [gr-qc]].

\bibitem{Sakstein:2015jca}
  J.~Sakstein and S.~Verner,
  ``Disformal Gravity Theories: A Jordan Frame Analysis,''
  Phys.\ Rev.\ D {\bf 92}, no. 12, 123005 (2015).
  [arXiv:1509.05679 [gr-qc]].

\bibitem{Domenech:2015tca}
  G.~Domenech, S.~Mukohyama, R.~Namba, A.~Naruko, R.~Saitou and Y.~Watanabe,
  ``Derivative-dependent metric transformation and physical degrees of freedom,''
  Phys.\ Rev.\ D {\bf 92}, no. 8, 084027 (2015).
  [arXiv:1507.05390 [hep-th]].
 
   \bibitem{Takahashi:2017zgr}
  K.~Takahashi, H.~Motohashi, T.~Suyama and T.~Kobayashi,
  ``General invertible transformation and physical degrees of freedom,''
  Phys.\ Rev.\ D {\bf 95}, no. 8, 084053 (2017).
  [arXiv:1702.01849 [gr-qc]].
 
\bibitem{Babichev:2019twf}
E.~Babichev, K.~Izumi, N.~Tanahashi and M.~Yamaguchi,
``Invertible field transformations with derivatives: necessary and sufficient conditions,''
[arXiv:1907.12333 [hep-th]].
 
\bibitem{Kamenshchik:2014waa} 
  A.~Y.~Kamenshchik and C.~F.~Steinwachs,
  ``Question of quantum equivalence between Jordan frame and Einstein frame,''
  Phys.\ Rev.\ D {\bf 91}, no. 8, 084033 (2015)
  [arXiv:1408.5769 [gr-qc]].

\bibitem{BenAchour:2020wiw}
J.~Ben Achour, H.~Liu and S.~Mukohyama,
``Hairy black holes in DHOST theories: Exploring disformal transformation as a solution-generating method,''
JCAP \textbf{02} (2020), 023
[arXiv:1910.11017 [gr-qc]].

\bibitem{Fewster:2010gm}
  C.~J.~Fewster and G.~J.~Galloway,
  ``Singularity theorems from weakened energy conditions,''
  Class.\ Quant.\ Grav.\  {\bf 28} (2011) 125009
  [arXiv:1012.6038 [gr-qc]]. 
  
\bibitem{Fewster:2019bjg}
C.~J.~Fewster and E.~A.~Kontou,
``A new derivation of singularity theorems with weakened energy hypotheses,''
Class. Quant. Grav. \textbf{37} (2020) no.6, 065010
[arXiv:1907.13604 [gr-qc]].

\bibitem{Galtsov:2020jnu}
D.~V.~Gal'tsov,
``Conformal and kinetic couplings as two Jordan frames of the same theory: Conformal and kinetic couplings,''
Eur. Phys. J. C \textbf{80} (2020) no.5, 443
[arXiv:2001.03221 [gr-qc]].

\bibitem{Rubakov:2014jja} 
  V.~A.~Rubakov,
  ``The Null Energy Condition and its violation,''
  Phys.\ Usp.\  {\bf 57}, 128 (2014)
  [Usp.\ Fiz.\ Nauk {\bf 184}, no. 2, 137 (2014)]
  [arXiv:1401.4024 [hep-th]].
  
   \bibitem{Mandal2019}
   Mandal, S.
   `` Revisiting Laws of Black Hole Mechanics and Violation of Null Energy Condition''.
    Journal of High Energy Physics, Gravitation and Cosmology, 5, 82-111 (2019)
   arXiv:1907.12333 [hep-th].
  
    \bibitem{Fisher}
   I.Z.  Fisher, 
  `` Scalar mesostatic field with regard for gravitational effects''
  Zh. Exp. Teor. Fiz., v. 18,  636-640  (1948), translated in arXiv:
gr-qc/9911008.

\bibitem{Bergmann:1957zza} 
  O.~Bergmann and R.~Leipnik,
  {\it Space-Time Structure of a Static Spherically Symmetric Scalar Field},
  Phys.\ Rev.\  {\bf 107} (1957) 1157.
  
  \bibitem{JNW}
  A.I. Janis, E.T. Newman, J. Winicour, 
  ``Reality of the Schwarzschild singularity''
  Phys. Rev. Lett. {\bf 20},  878 (1968).
     
\bibitem{Bronnikov:1973fh} 
  K.~A.~Bronnikov,
  ``Scalar-tensor theory and scalar charge,''
  Acta Phys.\ Polon.\ B {\bf 4}, 251 (1973). 
\bibitem{Abdolrahimi:2009dc}
S.~Abdolrahimi and A.~A.~Shoom,
``Analysis of the Fisher solution,''
Phys. Rev. D \textbf{81} (2010), 024035
[arXiv:0911.5380 [gr-qc]].

\end{thebibliography}
\end{document}